 \documentclass[prd,aps,
twocolumn,
nofootinbib,
floatfix,
superscriptaddress]{revtex4-2}
\usepackage{graphicx}
\usepackage{epsfig}
\usepackage{rotating}
\usepackage{amssymb}
\usepackage{amsfonts}
\usepackage{subfigure}
\usepackage{dsfont}
\usepackage{psfrag}
\usepackage{bbold}
\usepackage{xcolor}
\usepackage{multirow}
\usepackage{amsmath,euscript,array,mathrsfs}

\newcommand{\beq}{\begin{equation}}
\newcommand{\eeq}{\end{equation}}
\newcommand{\beqs}{\begin{eqnarray}}
\newcommand{\eeqs}{\end{eqnarray}}

\newcommand{\Tr}{{\rm Tr}}

\def\hbar{\hspace{0pt}\raisebox{1pt}{$-$} \hspace{-7pt} h}

\newcommand{\be}{\begin{equation}}
\newcommand{\ee}{\end{equation}}

\newcommand{\bea}{\begin{eqnarray}}
\newcommand{\eea}{\end{eqnarray}}

\def\lbldef#1#2{\expandafter\gdef\csname #1\endcsname {#2}}

\def\href#1#2{#2}

\newcommand{\ber}{\begin{eqnarray}}
\newcommand{\eer}{\end{eqnarray}}

\newcommand{\beqar}{\begin{eqnarray}}

\newcommand{\eeqar}{\end{eqnarray}}

\newcommand{\dsl}
  {\kern.06em\hbox{\raise.15ex\hbox{$/$}\kern-.56em\hbox{$\partial$}}}

\newcommand{\eeqarr}{\end{eqnarray}}
\newcommand{\ZZ}{{\rm \kern 0.275em Z \kern -0.92em Z}\;}

\def\CC{{\mathchoice
{\rm C\mkern-8mu\vrule height1.45ex depth-.05ex
width.05em\mkern9mu\kern-.05em}
{\rm C\mkern-8mu\vrule height1.45ex depth-.05ex
width.05em\mkern9mu\kern-.05em}
{\rm C\mkern-8mu\vrule height1ex depth-.07ex
width.035em\mkern9mu\kern-.035em}
{\rm C\mkern-8mu\vrule height.65ex depth-.1ex
width.025em\mkern8mu\kern-.025em}}}

\def\RR{{\rm I\kern-1.6pt {\rm R}}}

\def\ZZ{{\rm Z}\kern-3.8pt {\rm Z} \kern2pt}
\def\IB{\relax{\rm I\kern-.18em B}}
\def\ID{\relax{\rm I\kern-.18em D}}
\def\II{\relax{\rm I\kern-.18em I}}
\def\IP{\relax{\rm I\kern-.18em P}}

\newcommand{\bear}{\begin{eqnarray}}
\newcommand{\eear}{\end{eqnarray}}

\def\cl{{closed}}

\def\6{\partial}

\def\bea{\begin{eqnarray}}
\def\eea{\end{eqnarray}}

\def\beqx{\begin{displaymath}}
\def\eeqx{\end{displaymath}}

\newcommand{\bmat}{\left(\begin{array}}
\newcommand{\emat}{\end{array}\right)}

\def\cl{{\cal L}}

\def\bo{{\raise-.3ex\hbox{\large$\Box$}}}               
\def\face{{\raise.2ex\hbox{$\displaystyle \bigodot$}\mskip-2.2mu \llap {$\ddot
        \smile$}}}                                   
\def\>{\rangle}                                      
\def\<{\langle}                                      

\def\leftrightarrowfill{$\mathsurround=0pt \mathord\leftarrow \mkern-6mu
        \cleaders\hbox{$\mkern-2mu \mathord- \mkern-2mu$}\hfill
        \mkern-6mu \mathord\rightarrow$}        
\def\dvec#1{\vbox{\ialign{##\crcr
        \leftrightarrowfill\crcr\noalign{\kern-1pt\nointerlineskip}
        $\hfil\displaystyle{#1}\hfil$\crcr}}}           
\def\Tr{{\rm Tr \,}}                                    

\def\-{\hphantom{-}}

\begin{document}

\title{Composite two-Higgs doublet model from dilaton effective field theory}

\author{Thomas Appelquist}
\affiliation{Department of Physics, Sloane Laboratory, Yale University, New Haven, Connecticut 06520, USA}
\author{James Ingoldby}
\affiliation{Abdus Salam International Centre for Theoretical Physics, Strada Costiera 11, 34151, Trieste, Italy}
\author{Maurizio Piai}
\affiliation{Department of Physics, Faculty  of Science and Engineering,
Swansea University, Singleton Park, SA2 8PP, Swansea, Wales, UK}



\begin{abstract}
We construct a composite two-Higgs-doublet model (2HDM) within the context of dilaton effective field theory. This EFT describes the particle spectrum observed in lattice simulations of a near-conformal $SU(3)$ gauge field theory. A second Higgs doublet is naturally accommodated within the EFT, used previously to construct a one-Higgs-doublet model. We explore the interplay of phenomenological and EFT requirements, supplemented by information from numerical lattice studies of the $SU(3)$ gauge theory with eight fundamental (Dirac) fermions. We build a viable model with moderate parameter tuning. In the relevant portion of parameter space, we match the model to a conventional general description of 2HDMs, and comment on the role of custodial symmetry (breaking) in the scalar potential. Contributions to the precision electroweak parameter $T$ provide a possible explanation for the recent measurement of the W-boson mass by CDF II.	
\end{abstract}

\maketitle


\section{Introduction}
\label{sec:Intro}

Theoretical work following the discovery of the Higgs boson~\cite{Aad:2012tfa,Chatrchyan:2012xdj} is essential to identify signals of new physics detectable in the next generation of experimental searches. Much interest has focused on composite Higgs models (CHMs)~\cite{Kaplan:1983fs,Georgi:1984af,Dugan:1984hq} (see also Refs.~\cite{Panico:2015jxa,Panico:2015jxa,Witzel:2019jbe,Cacciapaglia:2020kgq,Ferretti:2016upr,Cacciapaglia:2019bqz} and references therein). Here the Higgs fields appear within the low energy EFT description of pseudo-Nambu-Goldstone bosons (PNGBs), themselves originating from a new, strongly coupled field theory. Lattice studies have begun to provide non-perturbative information about the dynamics of such strongly coupled systems, and their spectroscopy~\cite{Hietanen:2014xca,Detmold:2014kba,Arthur:2016dir,Arthur:2016ozw,Pica:2016zst,Lee:2017uvl,Drach:2017btk,Ayyar:2017qdf,Ayyar:2018zuk,Ayyar:2018ppa,Ayyar:2018glg,Cossu:2019hse,Bennett:2017kga,Bennett:2019jzz,Bennett:2019cxd,Drach:2020wux,Drach:2021uhl,DelDebbio:2021xlv,Bennett:2022yfa}.

A surprising feature emerged in the lattice studies of an $SU(3)$ gauge theory with  $N_f = 8$ fundamental Dirac fermions~\cite{Aoki:2014oha,Appelquist:2016viq,Aoki:2016wnc,Gasbarro:2017fmi,Appelquist:2018yqe}, or $N_f = 2$ Dirac fermions in the symmetric representation~\cite{Fodor:2012ty,Fodor:2015vwa,Fodor:2016pls,Fodor:2017nlp,Fodor:2019vmw,Fodor:2020niv}: an anomalously light composite state with vacuum quantum numbers. This particle can be interpreted as a dilaton (the PNGB associated with scale invariance), and it can be included in the low-energy EFT. The resulting dilaton EFT has a long history, dating back at least to early studies of dynamical symmetry breaking~\cite{Leung:1985sn,Bardeen:1985sm,Yamawaki:1985zg} (see also Refs.~\cite{Migdal:1982jp,Goldberger:2007zk,Coleman:1985rnk}), but the recent lattice measurements have led to a resurgence of interest~\cite{Matsuzaki:2013eva,Golterman:2016lsd,Kasai:2016ifi,Hansen:2016fri,Golterman:2016cdd,Appelquist:2017wcg,Appelquist:2017vyy,Golterman:2018mfm,Cata:2019edh,Appelquist:2019lgk,Golterman:2020tdq,Golterman:2020utm,LatticeStrongDynamicsLSD:2021gmp,Hasenfratz:2022qan}. The application of the dilaton EFT to lattice data yields a parameter measurement~\cite{Appelquist:2017wcg,Appelquist:2017vyy} consistent with a large anomalous dimension for the bilinear fermion condensate~\cite{Cohen:1988sq,Ryttov:2010iz,Ryttov:2016asb,Leung:1989hw}.

In a recent paper~\cite{Appelquist:2020bqj}, a viable CHM was built on the basis of the dilaton-EFT description of the $N_f = 8$ gauge theory. This CHM shares important features with earlier proposals~\cite{BuarqueFranzosi:2018eaj,Vecchi:2015fma,Ma:2015gra}, yet it exploits in a distinctive way the role of the dilaton field and existing lattice results. At the price of moderate fine tuning, only one Higgs boson, $h^0$, is accessible to the LHC. All additional composite states are heavy enough to evade even conservative bounds from direct~\cite{Sirunyan:2019vgj} and indirect~\cite{Aad:2019mbh} searches\footnote{We adopt $M_{\pi}=4$ TeV as an  indicative reference mass scale for the PNGBs. This scale could be lowered, and the fine tuning further softened, as some  experimental bounds from direct searches involve suppressed model-dependent couplings that enter certain production rates. See for instance Refs.~\cite{Belyaev:2016ftv,Cacciapaglia:2020vyf} for useful dedicated phenomenological studies.}.

Two-Higgs-doublet models predict the existence of two additional neutral scalar particles ($H^0$ and $A^0$) and two charged ones ($H^\pm$), each of which could be light enough to be within reach of future LHC direct searches. The treatment of generic 2HDMs with EFT principles leads to a rich phenomenology, but also to the daunting task of constraining many independent couplings (see, e.g., Refs.~\cite{Georgi:1978xz,Low:2020iua} and references therein). Bounds derived from perturbative unitarity and vacuum stability have been used for this purpose, but an organisational principle is naturally provided in the case of 2HDMs which are CHMs~\cite{Mrazek:2011iu,DeCurtis:2021uqx}.

Recently, an improved measurement of the mass of the W boson by the CDF II collaboration has been published~\cite{CDF:2022hxs}, leading to some tension with precision measurements within the standard model. If confirmed, this measurement would suggest the existence of sources of custodial symmetry breaking in new physics. In particular, it suggests that the precision electroweak parameter $T$~\cite{Peskin:1990zt,Peskin:1991sw} lies in the range $0.1 - 0.3$, the precise value depending on the treatment of other experimental input and on the assumptions made about the combination with other precision parameters~\cite{Strumia:2022qkt,deBlas:2022hdk,Fan:2022yly,Lu:2022bgw,Paul:2022dds,DiLuzio:2022xns,Bagnaschi:2022whn,Asadi:2022xiy,Balkin:2022glu,Biekotter:2022abc,Almeida:2022lcs}. Several studies suggest that 2HDMs can accommodate these values~\cite{Babu:2022pdn,Bahl:2022xzi,Song:2022xts,Heo:2022dey,Ahn:2022xeq,Ghorbani:2022vtv,Lee:2022gyf,Abouabid:2022lpg,Fan:2022dck}.

In this paper, we reconsider the dilaton EFT we used to describe the $N_f = 8$ lattice data~\cite{Appelquist:2017wcg,Appelquist:2017vyy,Appelquist:2019lgk}, and that we deployed to build a minimal CHM~\cite{Appelquist:2020bqj}. We modify it in a simple way, by adding potential terms such that a composite 2HDM emerges at low energies, separated from the rest of the spectrum by a mass gap. We describe this construction in some detail, use lattice measurements to constrain the model, discuss fine-tuning, and compute the spectrum of additional light scalars. We then match the resulting low energy theory to a conventional description of a generic 2HDM. We use the latter to estimate the 1-loop contribution to precision electroweak parameters. This demonstrates how the dilaton--EFT treatment of near-conformal theories provides a flexible model-building instrument with broad applicability. In this framework it is possible to explain the anomaly in the measurement of the W mass~\cite{CDF:2022hxs}.

\section{The Model}
\label{sec:Model}

\begin{table}[t]
	\vspace{10pt}
	\centering
	\renewcommand\arraystretch{1.2}
	\begin{tabular}{| c |  c  c  c | c |}
		\hline\hline
		Fermion
		& $SU(2)_L$ & $U(1)_Y$ & $SU(3)_c$& $SU(3)$\\
		\hline
		$L_\alpha$  & 2 & 0 & 1 & 3\\
		$R_{1,2}$  & 1 & $\begin{pmatrix} 1/2\\ -1/2 \end{pmatrix}$ & 1 & 3 \\
		$\cal T$  & 1 & 2/3 & 3 & 3  \\
		$\cal S$  & 1 & 0  & 1 & 3 \\
		\hline\hline
	\end{tabular}
	\caption{Quantum number assignments of the Dirac fermions. $SU(2)_L \times U(1)_Y \times SU(3)_c$ 
		is the SM gauge group, while $SU(3)$ is the strongly coupled gauge symmetry. 
		The $SU(2)_L$ index is $\alpha=1,\,2$.
		The fermions  $R_{1,2}$ form
		a fundamental representation of the global $SU(2)_R$. See also
		Ref.~\cite{Vecchi:2015fma}.}\label{Fig:fields}
\end{table}

We summarise in Table~\ref{Fig:fields}
the standard--model (SM) quantum numbers of the fermions of the $N_f=8$, $SU(3)$ gauge theory, with the same assignments as in Ref.~\cite{Appelquist:2020bqj}. The theory confines, and its approximate, global $SU(8)_1\times SU(8)_2$ symmetry is broken by the condensate to the diagonal $SU(8)_V$ subgroup.
Long distance physics is captured by an EFT containing only the fields associated with the dilaton and the $63$ PNGBs.
The  Lagrangian density is
the following:
\beqs
{\cal L} &=& \frac{1}{2}\partial_{\mu}\chi\partial^{\mu}\chi -V_{\Delta} + {\cal L}_{\pi}+{\cal L}_M + {\cal L}_Y  - V_{f}\,.
\label{Eq:L}
\eeqs
The first four terms of Eq.~(\ref{Eq:L}) provide the dilaton--EFT description of the strong dynamics, considered in isolation. The other terms describe additional interactions arising from underlying couplings of the fermions to the SM fields and to each other.

The field $\chi$ describes the dilaton. The (dilatation-symmetry-breaking) interactions are encoded in the potential $V_{\Delta}$, which  depends only on $\chi$.
This potential originates from the intrinsic breaking of scale invariance arising from the strong dynamics of the gauge theory.
We use the form for the potential discussed in Ref.~\cite{Appelquist:2019lgk}, which assumes 
that the breaking of dilatation symmetry is due to the coupling of an operator of dimension $\Delta$, without 
further committing to a choice of $\Delta$.
The field $\chi$ appears also in all other terms of the Lagrangian density, reflecting its role as a conformal compensator~\cite{Coleman:1985rnk}.

The $SU(8)_1\times SU(8)_2/SU(8)_V$ coset is described in the EFT by the
matrix-valued field $\Sigma$, subject to the non-linear 
constraint $\Sigma^{\dagger}\Sigma=\mathbb{1}_8$.
The kinetic terms for the $63$ PNGBs derive from
\beqs
{\cal L}_{\pi}&=&
\frac{F_{\pi}^2}{4}\left(\frac{\chi}{F_d}\right)^2
\Tr \left[\frac{}{}D_{\mu}\Sigma (D^{\mu}\Sigma)^{\dagger}\right]\,,
\eeqs
where $F_{\pi}$ and $F_{d}$ are parameters that control the scale of spontaneous breaking of 
global and dilatation symmetry, respectively. The nonlinear constraint leads to $\Sigma=\exp\left[\frac{2i}{F_{\pi}}\sum_A\pi^At^A\right]$, where $\pi^A$ are the $63$ real PNGBs, and $t^A$ the $SU(8)_V$ generators, normalised with $\Tr t^At^B=\frac{1}{2}\delta^{AB}$. The presence of covariant derivatives reflects the (weak) gauging of the $SU(2)_L\times U(1)_Y\times SU(3)_c$ 
subgroup of the unbroken $SU(8)_V$ global symmetry.
The corresponding gauge fields have a standard Lagrangian density, which we leave implicit.

Explicit breaking of the global symmetry is
also due to a universal mass term for the fundamental fermions.
Its effects are captured in the EFT by 
\beqs
{\cal L}_M&=& \frac{M_{\pi}^2F_{\pi}^2}{4}\left(\frac{\chi}{F_d}\right)^y \Tr\left[\Sigma + \Sigma^{\dagger}\right]\,,
\label{eq:lm}
\eeqs
where $M_{\pi}^2$ is the mass of the PNGBs, while $y$ is the 
parameter mentioned in the introduction, related to the scaling dimension of the chiral condensate. Fits to the lattice data indicate a value close to $y\simeq2$~\cite{Appelquist:2017wcg,Appelquist:2017vyy,Appelquist:2019lgk}. This term also contributes to the dilaton potential, modifying both its vacuum expectation value (VEV) and mass.

To describe the presence of two Higgs doublets within the PNGB multiplet,
in the basis defined by the field content in Table~\ref{Fig:fields}, we introduce the following matrices:
\begin{align}
	P_\alpha=\begin{pmatrix}
		\tilde{P}_\alpha & \mathbb{0}_{4}\\
		\mathbb{0}_{4} & \mathbb{0}_{4}
	\end{pmatrix},
	\label{eq:P}
\end{align}
with
\begin{align} 
	\tilde{P}_1=\frac{1}{2}\begin{pmatrix}
		0 & 0 & 1 & 0  \\
		0 & 0 & 0 & 0  \\
		0 & 0 & 0 & 0  \\
		0 & -1 & 0 & 0  \\
	\end{pmatrix},\quad
	\tilde{P}_2=\frac{1}{2}\begin{pmatrix}
		0 & 0 & 0 & 0  \\
		0 & 0 & 1 & 0  \\
		0 & 0 & 0 & 0  \\
		1 & 0 & 0 & 0  \\
	\end{pmatrix}.\label{eq:Pmatrix}
\end{align}
and 
\begin{align}
	R_\alpha=\begin{pmatrix}
		\tilde{R}_\alpha & \mathbb{0}_{4}\\
		\mathbb{0}_{4} & \mathbb{0}_{4}
	\end{pmatrix},
	\label{eq:R}
\end{align}
with
\begin{align} 
	\tilde{R}_1=\frac{1}{2}\begin{pmatrix}
		0 & 0 & i & 0  \\
		0 & 0 & 0 & 0  \\
		0 & 0 & 0 & 0  \\
		0 & i & 0 & 0  \\
	\end{pmatrix},\quad
	\tilde{R}_2=\frac{1}{2}\begin{pmatrix}
		0 & 0 & 0 & 0  \\
		0 & 0 & i & 0  \\
		0 & 0 & 0 & 0  \\
		-i & 0 & 0 & 0  \\
	\end{pmatrix}.\label{eq:Rmatrix}
\end{align}
The combinations $\Tr\left[P_{\alpha}\Sigma\right]$ and
$\Tr\left[R_{\alpha}\Sigma\right]$
transform as two Higgs doublets $H^1_{\alpha}$ and $H^2_{\alpha}$, respectively,
with identical quantum numbers $(2,-1/2,1)$,
under the SM group $SU(2)_L\times U(1)_Y\times SU(3)_c$. 
The conjugate Higgs doublets are obtained with the second Pauli matrix $\tau^2$, as 
$\tilde{H}_{\alpha}^1 \sim -i(\tau^2)_{\alpha}^{\,\,\,\,\beta}\Tr\left[P_{\beta}^\dagger\Sigma\right]$ 
and $\tilde{H}_{\alpha}^2 \sim -i(\tau^2)_{\alpha}^{\,\,\,\,\beta}\Tr\left[R_{\beta}^\dagger\Sigma\right]$.

New underlying interactions, extending the $SU(3)$ gauge theory,
are assumed to generate Yukawa-like couplings to the SM fermions,
breaking the $SU(8)_1 \times SU(8)_2$ symmetry. We 
write the resulting couplings to the third-family quarks in the simple form:
\begin{multline}
	{\cal L}_Y=y_t F_{\pi} \left(\frac{\chi}{F_d}\right)^{z}
	\bar{Q}_L^{\alpha}\left(\Tr\left[P_{\alpha} \Sigma\right]\right) t_R\\
	+y_b F_{\pi} \left(\frac{\chi}{F_d}\right)^{z}\bar{Q}_L^{\alpha}\left(-i(\tau^2)_{\alpha}^{\,\,\,\,\beta} 
	\Tr\left[P_{\beta}^\dagger \Sigma\right]\right) b_R+\text{h.c}\,
	,\label{eq:ly}
\end{multline}
where $z$ is a scaling dimension and $y_t$ and $y_b$ are the top and bottom Yukawa couplings. Replacing them by complex $3 \times 3$ matrices would allow for a straightforward extension
to three SM families. We introduce a $Z_2$ symmetry, under which only $H^2_\alpha\sim\Tr\left[R_\alpha\Sigma\right]$ is odd. As a result, we couple to the fermions only one of the doublets, corresponding to $H^1_{\alpha}$. Fermion mass hierarchies and mixing are entirely attributed to the hierarchies in the $y_i$ couplings. This choice implements a GIM mechanism that protects the model from excessive flavor-changing neutral current processes.

It is natural to expect that the underlying interactions responsible for generating $\cl_Y$ lead also to new contributions to the potential of the Higgs doublets. To accommodate the two doublets, we allow for terms involving both $P_{\alpha}$ and $R_{\alpha}$. We assume the dominant terms to be quadratic in these structures and write this new potential in the form
\begin{widetext}
	\begin{align}
		V_f =  -\frac{C}{4}\left(\frac{\chi}{F_d}\right)^w\sum_{\alpha}&\left\{ (1+\kappa)\left[(1+\lambda)\left|\Tr\left[P_{\alpha}\Sigma\right]\right|^2+(1-\lambda)\left|\Tr\left[P_{\alpha}^\dagger\Sigma\right]\right|^2\right]\right.\nonumber\\
		&\left.+(1-\kappa)\left[(1+\lambda)\left|\Tr\left[R_{\alpha}\Sigma\right]\right|^2+(1-\lambda)\left|\Tr\left[R_{\alpha}^\dagger\Sigma\right]\right|^2\right]\right\}\,,
		\label{Eq:Vf}
	\end{align}
\end{widetext}
where the parameter $w$ is a scaling dimension. We have omitted terms that mix operators built with $P_\alpha$ and $R_{\alpha}$. This is motivated by invoking the $Z_2$ symmetry acting on $H^2_{\alpha}\sim\Tr[R_{\alpha}\Sigma]$. Because such mixing terms are absent, only $H^1_{\alpha}$ develops a non-vanishing VEV as long as $\kappa > 0$, making $H^2_{\alpha}$ an ``inert'' doublet.

The limits $\kappa\rightarrow 0$ and $\lambda\rightarrow 0$ are symmetry enhancement points: the former related to a $U(1)_X$ symmetry making the two Higgs doublets degenerate in mass, the latter to  custodial $SU(2)_L\times SU(2)_R$ symmetry.
We assume that $C \geq 0$, hence triggering electroweak symmetry breaking (EWSB), 
and will discuss its magnitude in the next section. The potential $V_t$ of Ref.~\cite{Appelquist:2020bqj} is recovered by setting $C = C_t$, $\lambda = 1$, and $\kappa = 1$.

For completeness, we show explicitly how the aforementioned $U(1)_X$ symmetry acts on the PNGB fields:
\begin{equation}
	\Sigma\rightarrow X \Sigma X^\dagger,
	\label{eq:X1}
\end{equation}
where $X X^\dagger=\mathbb{1}$ and
\begin{equation}
	{X} = \text{diag}\left(e^{i\frac{\phi}{2}},\,e^{i\frac{\phi}{2}},\, e^{-i\frac{\phi}{2}},\, e^{-i\frac{\phi}{2}},1,1,1,1\right)\, .
	\label{eq:X2}
\end{equation}
Under the action of this symmetry, the two Higgs doublets are rotated into each other in the following way:
\begin{align*}
	\Tr\left[P_\alpha\Sigma\right] & \rightarrow \cos\phi \;\Tr\left[P_\alpha\Sigma\right] - \sin\phi \;\Tr\left[R_\alpha\Sigma\right]\, ,\\
	\Tr\left[R_\alpha\Sigma\right] & \rightarrow \cos\phi \;\Tr\left[R_\alpha\Sigma\right] + \sin\phi \;\Tr\left[P_\alpha\Sigma\right]\, .
\end{align*}

\section{The vacuum}
\label{sec:Vacuum}

The vacuum structure is determined by minimising the potential derived form Eq.~(\ref{Eq:L}).
When $\kappa>0$, only the doublet $H^1_{\alpha}$ 
acquires a nonzero value vacuum expectation value. We parametrise the EWSB vacuum value of the $\Sigma$ field as
\beqs
\langle \Sigma \rangle&\equiv &\exp\left[i\theta \left(
\begin{array}{ccc}
	\mathbb{0}_{2\times 2} & -i \mathbb{1}_{2} & \mathbb{0}_{2\times 4}\cr
	i \mathbb{1}_{2} &\mathbb{0}_{2\times 2} & \mathbb{0}_{2\times 4}\cr
	\mathbb{0}_{4 \times 2} &\mathbb{0}_{2\times 2}& \mathbb{0}_{4}
\end{array}\right)\right],
\eeqs
with $0\leq \theta \leq \pi$. The potential to be minimised to determine $\theta$ and $\chi$ is the following:
\beqs
W&\equiv V_{\Delta}(\chi)&-{\cal L}_{M}(\chi,\theta)+V_f(\chi,\theta)\nonumber\\
&=V_{\Delta}(\chi)&-2M_{\pi}^2F_{\pi}^2 \left(\frac{\chi}{F_d}\right)^y(1+\cos\theta)\\
&  & -\frac{C}{2}\left(\frac{\chi}{F_d}\right)^w\left(1+\kappa\right)\sin^2\theta\nonumber\,.
\eeqs
This potential does not depend on the parameter $\lambda$ that breaks custodial symmetry,
but the spectrum does, as we shall see explicitly.
As in Ref.~\cite{Appelquist:2020bqj}, we define $\langle \chi \rangle \equiv F_d$.
This potential is minimised when $\theta$ takes the value
\beqs
\cos\theta &=&\frac{2 M_{\pi}^2F_{\pi}^2}{C\left(1+\kappa \right)}\,.
\label{Eq:tuning}
\eeqs

The EWSB vacuum is then given by
\beqs
v&\equiv&\sqrt{2} F_{\pi} \sin \theta\,\simeq\,246 \, {\rm GeV}\,,
\label{eq:vew}
\eeqs
providing only that $2M_{\pi}^2F_{\pi}^2< C(1+\kappa)$.
The masses of the SM gauge bosons are approximated by
\beqs
M_W^2&=&\frac{1}{2}g^2 F_{\pi}^2 \sin^2\theta\,,\\
M_Z^2&=&\frac{1}{2}(g^2+g^{\prime\,2}) F_{\pi}^2 \sin^2\theta\,,
\eeqs
for the $W$ and $Z$ bosons, respectively.
For the SM fermions, the masses of the top and bottom quarks are
\beqs
m_{t,b}&=&\,y_{t,b}\frac{v}{\sqrt{2}}\,.
\eeqs

To ensure a hierarchy between the electroweak scale $v$ and the confinement scale of the gauge theory, $\theta\ll1$ is necessary. This is achieved by a single tuning, of the parameter combination $C(1 + \kappa)$ against the combination $M_{\pi}^{2} F_{\pi}^{2}$. The factor $M_{\pi}^{2}$ is governed by the assigned mass of the fermions in the gauge theory. The tuning here, at roughly the $2\%$ level, is the same as we implemented in Ref.~\cite{Appelquist:2020bqj}, and we refer the reader to that paper for additional comments.

We note also that it is natural to take $\kappa<1$, which brings down the mass of the second Higgs doublet relative to those of the many other PNGBs in the dilaton EFT. The size of $\kappa$ is affected by symmetry--breaking effects coming from ${\cal L}_Y$. The first term in $V_f$, for example, receives contributions from a loop of top quarks, controlled by $y_t$ and regulated by the confinement scale of the $SU(3)$ gauge theory. This contribution is not larger than the required size of $C(1+\kappa)$, and supplements those of other, underlying interactions, which determine all the terms in $V_f$. Only taking $\kappa$ to extremely small values would introduce additional fine tuning.


\section{The spectrum}
\label{sec:Spectrum}

We study the mass spectrum of perturbations near the EWSB minimum of 
the potential. We treat as free parameters the quantities 
$\{M^2_\pi,\,M^2_d,\,F^2_\pi,\,F^2_d,\,y,\,v,\,m^2_h,\,\lambda,\,\kappa\}$.
Following Ref.~\cite{Appelquist:2020bqj},
we set $v$ and $m^2_h$ to their experimentally measured values and fix
$M_\pi=4$ TeV as a benchmark value. 
Lattice calculations in the $N_f=8$ gauge theory provide values of $M^2_\pi/F^2_\pi$ 
and $M^2_d/F^2_\pi$; we take the measurements with fermion mass  $am=0.005$, mid--range in Tables III and IV of Ref.~\cite{Appelquist:2018yqe}. 
We fix $y=2.06\pm0.05$ and $F^2_\pi/F^2_d=0.086\pm0.015$, taken from an earlier dilaton--EFT study~\cite{Appelquist:2019lgk}.

We identify a region of parameter space in which the particles associated with the
2HDM are parametrically lighter than the other spin-0 states.
To this end, we allow
the two new free parameters $\kappa$ and $\lambda$ to differ from 
the choices $\kappa=1=\lambda$ implicitly adopted in Ref.~\cite{Appelquist:2020bqj}. We take $\lambda$, which controls custodial symmetry breaking, to be $O(1)$, and positive since the spectrum depends on only $\lambda^2$.

The aforementioned requirements demand $\theta\simeq 0.17$,
which is obtained by tuning $C$ in Eq.~(\ref{Eq:tuning}).
Because $\theta$ is small, simple approximate expressions describe the masses of the Higgs states. 
It was observed in Ref.~\cite{Appelquist:2020bqj} that mixing between one PNGB and the dilaton state leads to a simple expression for $m_h^{2}/v^2$:
\beq
\frac{m^2_h}{v^2}\simeq\frac{M^2_\pi}{2F^2_\pi}\left(1-\frac{2F^2_\pi M^2_\pi(y-w)^2}{F^2_d M^2_d}\right)\,,
\label{Eq:mhiggs}
\eeq
and this relation holds also for the present model.
The dilaton plays a key role in making this ratio realistically small. The combination of parameters in the parentheses in Eq.~(\ref{Eq:mhiggs}) provides the necessary cancellation, lowering the Higgs-mass-to-VEV ratio.

\begin{center}
	\begin{figure}[t]
		\includegraphics[width=0.95\columnwidth]{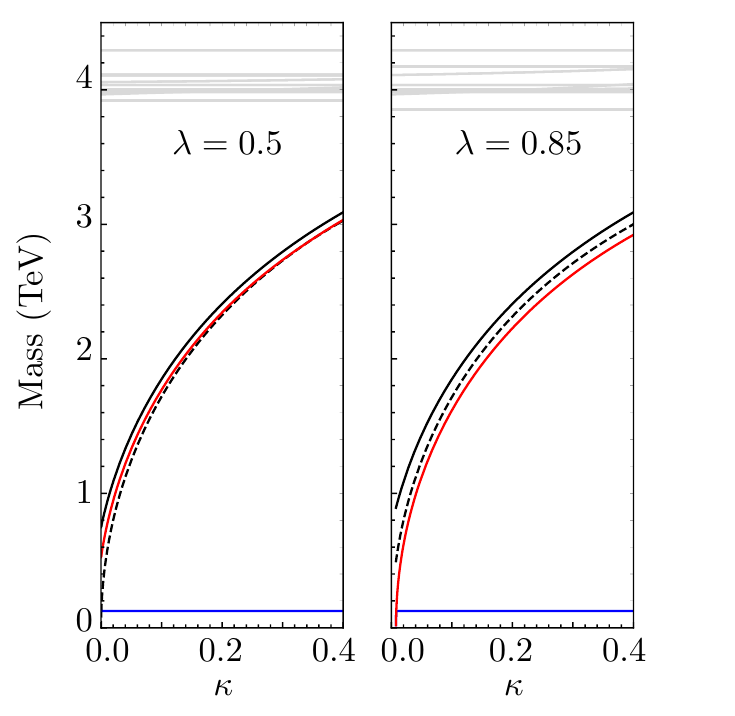}
		\caption{Masses for the composite states included in the EFT as a function of $\kappa$, taking $\lambda=0.5$ ($\lambda=0.85$) for the left (right) panel. Choices for the other model parameters are explained in the text. The blue lines refers to the state corresponding to the SM Higgs. The black solid lines refers to the $A^0$, the reds to the $H^\pm$ and the black dashed lines to the $H^0$ mass. The pale gray lines at the top of the spectrum indicate the masses of the remaining composite spinless states.}
		\label{fig:spectrum}
	\end{figure}
\end{center}

For illustration purposes, we take $\lambda=0.5$ and 0.85,
compute the masses of the dilaton and PNGB states, and show the results in Fig.~\ref{fig:spectrum},
varying the EFT parameter $\kappa>0$. 
As $\kappa$ is reduced, the masses of the four states which 
make up the second Higgs doublet split apart from the rest of the spectrum, represented 
by the pale gray lines sitting around $4$ TeV. For small $\kappa$, the low energy spectrum 
becomes that of a 2HDM, as desired. For instance, when $\kappa=0.1$ and $\lambda=0.85$, the masses of the four scalars associated with $H^2_\alpha$ lie around 1.6 TeV. Similar results apply for other generic values of $\lambda$.

The $U(1)_X$ symmetry shown in Eqs.~(\ref{eq:X1}) and (\ref{eq:X2})
interchanges the two Higgs doublets.
Since $\kappa$ breaks this symmetry,  when $\kappa$ is small the second Higgs doublet and SM Higgs states 
come closer together.  Taking $\kappa$ to zero would restore $U(1)_X$ in the potential 
$V_f$ of Eq.~(\ref{Eq:Vf}).

To provide simple analytic expressions for the second Higgs doublet masses, we take $\kappa\sim\theta^2\ll1$, a magnitude consistent with the phenomenological discussion to come. For the charged states $H^{\pm}$, we find
\beq
M^2_{H^\pm}= 2M^2_\pi\kappa+M^2_\pi\theta^2(1-2\lambda^2),
\label{eq:hp}
\eeq
plus terms higher order in $\kappa$ and $\theta^2$. We see also that for $|\lambda|>1/\sqrt{2}$, $M^2_{H^\pm}$ becomes negative for extreme values of $\kappa$. We avoid this region of parameter space in the following. The masses of the neutral CP-even Higgs state, $H^0$, and the 
CP-odd state, $A^0$, are given by
\beq
M^2_{H^0} = 2M^2_\pi \kappa,
\label{eq:mh0}
\eeq
and
\beq
M^2_{A^0}= 2M^2_\pi\kappa+M^2_\pi\theta^2.
\label{eq:a0}
\eeq
plus terms higher order in $\kappa$ and $\theta^2$. In the small $\lambda$ limit,
where custodial symmetry is restored, $H^{\pm}$ and $A^0$ become degenerate.
For $\lambda\simeq 1/\sqrt{2}$ one finds approximate degeneracy 
of the $H^0$ and $H^{\pm}$ states, up to subleading corrections in the small parameters $\kappa$ and $\theta$.

By choosing a small value of $\kappa$, the four additional particles of the second Higgs doublet can be light enough that they separate from the rest of the EFT spectrum. Because $\kappa$ is a parameter that breaks the approximate $U(1)_X$ global symmetry (broken only by $\cl_Y$), only a moderate amount of additional tuning, beyond that of Ref.~\cite{Appelquist:2020bqj}, is needed in this process. Also, since these results do not depend sensitively on the value of $\lambda$, the construction we have outlined may lead to spectra exhibiting custodial symmetry, or violating it, affecting the electroweak precision parameter $T$.



\section{Two--Higgs Doublet EFT}
\label{sec:eft}

With $\kappa$ small, the eight states comprising the two Higgs doublets are much lighter than the remaining states. An EFT for the two light doublets includes a potential with operators of dimension $d\leq4$. The most general form of the potential, in the conventions of Ref.~\cite{Low:2020iua}, is
\begin{align}
V&= m_{11}^2H^\dagger_1H_1+m_{22}^2H^\dagger_2H_2-\left(m_{12}^2H^\dagger_1H_2+\text{h.c.}\right)\nonumber\\
&+\frac{\beta_1}{2}\left(H_1^\dagger H_1\right)^2+\frac{\beta_2}{2}\left(H_2^\dagger H_2\right)^2+\beta_3\left(H_1^\dagger H_1\right)\left(H_2^\dagger H_2\right)\nonumber\\
&+\beta_4\left(H_1^\dagger H_2\right)\left(H_2^\dagger H_1\right)+\left\{\frac{\beta_5}{2}\left(H_1^\dagger H_2\right)^2+\text{h.c.}\right\}\nonumber\\
&+\left\{\left[\beta_6\left(H^\dagger_1 H_1\right)+\beta_7\left(H^\dagger_2 H_2\right)\right]\left(H^\dagger_1H_2\right)+\text{h.c.}\right\}\, .
\label{eq:Veft}
\end{align}
Imposing a $Z_2$ symmetry with respect to the second Higgs doublet sets
$m^2_{12}=\beta_6=\beta_7=0$. The relation of the other parameters in Eq.~(\ref{eq:Veft}) to the free parameters in our theory is shown in Table~\ref{tab:eftparams}, where we have kept only leading terms in the limit $\kappa\sim\theta^2\ll1$. The expressions for $m^2_{11}$ and $\beta_1$ lead directly to Eq.~(\ref{Eq:mhiggs}). Similarly, the expressions for $m^2_{22}$, $\beta_2$, $\beta_3$, $\beta_4$, and $\beta_5$ lead to the mass formulae Eqs.~(\ref{eq:hp})--(\ref{eq:a0}). We notice that couplings $\beta_1$ and $\beta_2$ are $O(m^2_h/v^2)\sim1/4$, typical of the electroweak theory whereas $\beta_3,\,\beta_4,\,\beta_5$ are larger. They are of order $O(M^2_\pi/F^2_\pi)\sim 16$, using lattice data~\cite{Appelquist:2018yqe}. The second Higgs doublet $H_2$ does not acquire a VEV or couple to the SM fermions. The model therefore reduces at low energies to an inert two–Higgs doublet model.

\begin{table}
	\centering
	\renewcommand\arraystretch{2.0}  
	\addtolength{\tabcolsep}{1.5 pt} 
	\begin{tabular}{|c|c|}
		\toprule
		\multirow{2}{2cm}{\centering 2HDM EFT Parameter} &  \multirow{2}{2cm}{\centering Dilaton EFT Expression} \\
		& \\
		\hline
		$m^2_{11}$ & $-m^2_h/2$ \\
		$m^2_{22}$ & $2M^2_\pi\kappa-m^2_h/2$ \\
		$\beta_1$ & $\frac{m^2_h}{v^2}$ \\
		$\beta_2$ & $\frac{m^2_h}{v^2}$ \\
		$\beta_3$ & $\frac{M^2_\pi}{F^2_\pi}\left(1-2\lambda^2\right)+\frac{m^2_h}{v^2}$  \\
		$\beta_4$ & $-\frac{M^2_\pi}{2F^2_\pi}\left(1-4\lambda^2\right)$\\
		$\beta_5$ & $-\frac{M^2_\pi}{2F^2_\pi}$\\
		\hline\hline
	\end{tabular}
	\caption{The parameters in the potential of the two-Higgs doublet model EFT as shown in Eq.~(\ref{eq:Veft}), expressed in terms of quantities in the dilaton EFT. We have expanded these relations in powers of the small quantities $\kappa,\,\theta^2$ and kept only the lowest--order nonvanishing terms, for simplicity. The expressions for $m^2_{11}$ and $m^2_{22}$ are truncated to include only terms linear in $\kappa$ and $\theta^2$. The expressions for couplings $\beta_i$ include only the zeroth order terms in $\kappa$ and $\theta^2$. We have simplified the expressions using Eqs.~(\ref{eq:vew}) and (\ref{Eq:mhiggs}).}
	\label{tab:eftparams}
\end{table}

\subsection{Electroweak Precision Observables}

We now turn our attention to the contribution of the new composite sector to the $S$ and $T$ electroweak precision parameters \cite{Peskin:1990zt,Peskin:1991sw}. The contribution from one--loop diagrams calculated using the $d\le 4$ potential shown in Eq.~(\ref{eq:Veft}) dominates over tree level contributions from higher dimension operators. 

The $T$ parameter is given by~\cite{Haber:1993wf,Haber:1999zh}\footnote{To apply the general formula for $T$ provided in appendix A of Ref.~\cite{Haber:1999zh} to our inert two--Higgs doublet model, we set $\sin(\beta-\alpha)=1$, using the notation of the reference.}:
\begin{multline}
T=\frac{1}{16\pi s^2_W m^2_W}\left({\cal F}(M_{H^\pm},M_{A^0})+{\cal F}(M_{H^\pm},M_{H^0})\right.\\ \left.-{\cal F}(M_{A^0},M_{H^0})\right),
\label{eq:tpar}
\end{multline}
where $m_W$ is the mass of the W boson and $s^2_W=\sin^2\theta_W$. We use the values $m_W=80.4$ GeV and $s^2_W=0.23$. The function ${\cal F}$ is defined as
\beq
{\cal F}(m_1,m_2) \equiv \frac{1}{2}\left(m^2_1+m^2_2\right) - \frac{m^2_1m^2_2}{m^2_1-m^2_2}\log\left(\frac{m^2_1}{m^2_2}\right).
\label{eq:fdef}
\eeq

\begin{center}
	\begin{figure}[h]
		\includegraphics[width=0.9\columnwidth]{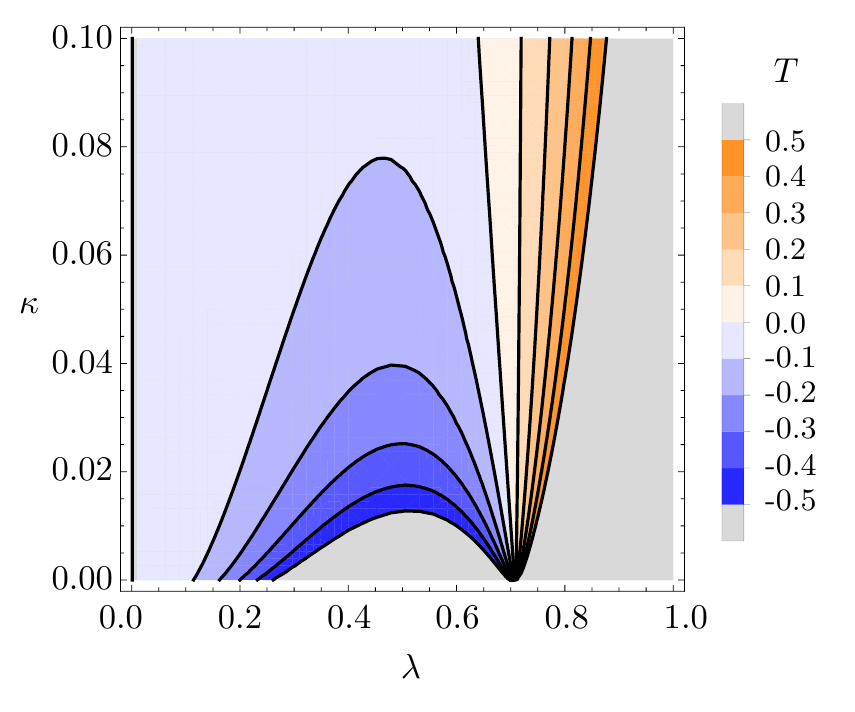}
		\caption{Contours of the electroweak $T$ parameter in the $\lambda$, $\kappa$ plane. The $T$ parameter is an even function of $\lambda$, so we show only the $\lambda>0$ region. In the areas shaded gray, $|T|>0.5$ making this parameter space disfavored by experiment.}
		\label{fig:tpar}
	\end{figure}
\end{center}

In Fig.~\ref{fig:tpar}, we use Eqs.~(\ref{eq:tpar}) and (\ref{eq:fdef}) to evaluate $T$ for a range of values of the parameters $\lambda$ and $\kappa$. The remaining model parameters are fixed using the experimental values for $m^2_h$ and $v$, and using lattice data as explained earlier. The masses of the second Higgs doublet states are required inputs for the calculation and we determine them using the potential of the full dilaton EFT at the classical level. The model allows $T$ to take both positive and negative values, and in general the magnitude of $T$ increases as $\kappa$ is reduced and the states of the second Higgs doublet become light. When $\lambda=0$, the scalar sector contains no sources of custodial symmetry violation and therefore $T$ vanishes. 

Fig.~\ref{fig:tpar} shows that $T$ changes sign near the threshold $\lambda\simeq1/\sqrt{2}$. This feature may be explained by considering the intermediate regime where $m^2_h\ll M^2_{H^0,\,A^0,\,H^\pm}\ll M^2_\pi$. In this case, the two--Higgs doublet EFT may still be used but Eq.~(\ref{eq:tpar}) can be simplified to yield \cite{Haber:2010bw}
\beq
T\simeq \frac{\left(M^2_{H^\pm}-M^2_{A^0}\right)\left(M^2_{H^\pm}-M^2_{H^0}\right)}{48\pi m^2_W s^2_W M^2_{A^0}}\,.
\eeq
In this approximate formula, the sign of $T$ changes when the mass hierarchy within the second Higgs doublet changes. By comparing Eq.~(\ref{eq:hp}) with (\ref{eq:mh0}), we see that the mass difference $M^2_{H^\pm}-M^2_{H^0}$ changes sign around $\lambda\simeq1/\sqrt{2}$.

The electroweak parameter $S$, when the two--Higgs doublet model is inert, is given by \cite{Haber:1993wf,Kanemura:2011sj}
\beq
S = \frac{1}{4\pi}\left({\cal F}^\prime(M_{A^0},M_{H^0})-{\cal F}^\prime(M_{H^\pm},M_{H^\pm})\right)\, ,
\eeq
where
\beqs
{\cal F}^\prime(m_{1},m_{2})\equiv-\frac{1}{3}&\left[\frac{4}{3}- \frac{m^2_1\log m^2_1-m^2_2\log m^2_2}{m^2_1-m^2_2}\right.\nonumber\\
&\left.-\frac{m^2_1+m^2_2}{(m_1^2-m^2_2)^2}{\cal F}(m_1,m_2)\right]\,.
\eeqs
We find that $|S|<0.02$ for all values of $\lambda$ explored in Fig.~\ref{fig:tpar} with $\kappa>0.01$. In this case, the contribution of our model to the $S$ parameter is negligible, unlike the contribution to $T$.

If the new, larger measurement for $m_W$ from the CDF collaboration \cite{CDF:2022hxs} is incorporated into a global electroweak precision fit, it favors positive values for $T$. Taking $S$ and the other electroweak precision parameters to be negligible, as is the case in our model, updated fits~\cite{Strumia:2022qkt,deBlas:2022hdk,Fan:2022yly,Paul:2022dds,DiLuzio:2022xns,Bagnaschi:2022whn,Asadi:2022xiy,Balkin:2022glu,Biekotter:2022abc,Almeida:2022lcs} suggest $T$ lying in the range $[0.1,\,0.3]$. In Fig.~\ref{fig:tpar}, we see that there is a portion of the $\{\kappa,\,\lambda\}$ parameter space for which these values of $T$ can be accommodated.

\section{Discussion}
\label{sec:Discussion}

We embedded into the dilaton-EFT framework a two-Higgs-doublet model (2HDM). As in Ref.~\cite{Appelquist:2020bqj}, we employed moderate fine tuning to yield a composite Higgs model, now with {\it two} Higgs doublets separated by a mass hierarchy from the other, heavier states in the dilaton EFT. The latter are heavy enough to escape both direct and indirect searches. The full dilaton EFT is partially UV completed by a near conformal gauge theory studied extensively on the lattice. We drew directly on these calculations.

We built an effective description of the resulting 2HDM, uncovering some features that are distinctive with respect to generic 2HDMs. For example, the quartic couplings $\beta_1$ and $\beta_2$ in Eq.~(\ref{eq:Veft}) are relatively small, of order unity, while the other scalar couplings are larger, of order $M_{\pi}^2/F_{\pi}^2$, as determined by the strong interactions of the near conformal gauge theory. Importantly, over significant portions of parameter space, the four additional scalars relative to the standard model form an inert doublet with near degenerate physical masses.

We explored the implications of these features for electroweak precision measurements, in particular with reference to custodial symmetry breaking. It will be interesting to understand whether other distinctive signatures emerge. Most notably, depending on parameter choices, the particles of the inert doublet could be light enough to appear in direct searches at future colliders. We leave these considerations for future dedicated studies.




\vspace{0.5cm}
\begin{acknowledgments}
	
The work of MP has been supported in part by the STFC Consolidated Grants ST/P00055X/1 and ST/T000813/1. MP has also received funding from the European Research Council (ERC) under the European Union's Horizon 2020 research and innovation programme under grant agreement No 813942.

\vspace{12pt}

{\bf Open Access Statement - } For the purpose of open access, the authors have applied a Creative Commons 
Attribution (CC BY) licence  to any Author Accepted Manuscript version arising.

\end{acknowledgments}

\end{document}